\documentclass[aps,epsfig,floats,twocolumn]{revtex4}
\usepackage{amssymb}
\usepackage{psfrag} 
\usepackage{latexsym} 
\usepackage{graphicx} 
\begin{document}
\unitlength 1 cm
\newcommand{\nn}{\nonumber}
\newcommand{\be}{\begin{equation}}
\newcommand{\ee}{\end{equation}}
\newcommand{\bearr}{\begin{eqnarray}}
\newcommand{\eearr}{\end{eqnarray}}
\newcommand{\bk}{\mathbf k}
\newcommand{\bq}{\mathbf q}
\newcommand{\bp}{\mathbf p}
\newcommand{\vk}{\vec k}
\newcommand{\vp}{\vec p}
\newcommand{\vq}{\vec q}
\newcommand{\vkp}{\vec {k'}}
\newcommand{\vpp}{\vec {p'}}
\newcommand{\vqp}{\vec {q'}}
\newcommand{\up}{\uparrow}
\newcommand{\down}{\downarrow}
\newcommand{\ns}{\normalsize}
\newcommand{\fns}{\footnotesize}
\newcommand{\cdag}{c^{\dagger}}

\title{Superconductivity in heavily vacant diamond}

\author{M. Alaei$^1$, S. Akbar Jafari$^1$ and H. Akbarzadeh$^1$}
\affiliation{
$^1$Department of Physics, Isfahan University of Technology, 
Isfahan 84156-83111, Iran}
\date{\today}

\begin{abstract}
Using first principle electronic structure calculations we investigated the
role of substitutional doping of B,N,P,Al and vacancies (${\cal V}$) in 
diamond (X$_{\alpha}$C$_{1-\alpha}$). In the heavy 
doping  regime, at about $\sim 1-6\%$ doping an impurity band appears in the 
mid gap. Increasing further the concentration of the impurity
substitution fills in the gap of the diamond host. 
Our first principle calculation indicates that in the case of vacancies, 
a clear single-band picture can be employed 
to write down an effective {\em one band} microscopic Hamiltonian, which 
can be used to further study various many-body and disorder effects in 
impurity band (super)conductors.
\end{abstract}

\maketitle

\section{introduction}
Diamond has a number of unique attributes that make it highly suited as a gem stone.
It is the hardest known material which can only be scratched by another diamond.
The thermal conductivity of diamond is the highest among all materials~\cite{Field,Davies}.
Irradiation of diamond by various particles (e.g. electrons, neutrons, $\alpha$ particles)
followed by annealing to repair damaged $sp^3$ bonds 
gives rise to fascinating colours of diamond, which are due to the so called colour 
centers~\cite{Collins}.

Diamond is also a material with semiconductor properties that are superior to 
Si, Ge, or GaAs, which are now commonly 
used. The use of diamond in electronic applications is not a new idea, 
but limitations in size and control of properties restricted the 
use of diamond to a few specialized applications. The vapor-phase 
synthesis of diamond, however, has facilitated serious interest in the 
development of diamond-based electronic devices. The process allows 
diamond films to be laid down over large areas. Both intrinsic and 
doped diamond films have a unique combination of extreme properties for high speed, 
high power and high temperature applications~\cite{Davies,Nebel,Lowrence}. 

Ekimov and coworkers~\cite{Ekimov} and subsequently 
Takano {\em et. al.}~\cite{Takano} used chemical vapor deposition (CVD)
to synthesized B-doped diamond. Doping diamond by low concentration of 
typically $10^{17}-10^{18}$~cm$^{-3}$ boron atoms gives rise
to acceptor level, rendering it to a $p$ type semiconductor~\cite{KalishBook}. 
Increasing the doping level to [B]/[C]$>5000$~ppm in the gas phase induces
metallic conductivity in diamond~\cite{Deneuville}. Further increasing the 
doping rate to the scale of $n> 10^{21}$~cm$^{-3}$, i.e. $\sim$ few~\%, makes it 
superconduct at low temperatures~\cite{Ekimov,Takano}. Increasing
the doping rate amounts to bringing the boron atoms closer, and allow them to 
overlap more effectively, which broadens the acceptor levels in to decent bands
of electrons~\cite{DanWu,Baskaran}, which are responsible for metallic and superconducting 
properties~\cite{Baskaran,KWLee,Pogorelov,Nakamura,Blase,Mukuda,Fukuyama}.

  Also some authors have investigated the doping of silicon with boron,
aluminum and phosphorus~\cite{BlaseAPL,BlaseNature}. Experiment has showed 
that the transition temperature for B-doped Si is $T_c\sim 0.35$~K. Therefore
doping C with boron gives more than an order of magnitude larger $T_c\sim 4$~K 
compared to B doped Si. Hence we choose the host material to be
carbon rather than any other element in the same group. 

   Now the question arises, what other elements can be doped into diamond 
in the regime of heavy ($\sim 1-10\%$) doping which can possibly lead to
higher superconducting $T_c$. In the example of high temperature cuprate
superconductors (HTSC)~\cite{Anderson}, an effective one-band model 
for the so called Zhang-Rice singlet~\cite{ZRS} can be written down
in terms of the hole states of the O$2p$ and Cu$3d$. In $X_{\alpha}C_{1-\alpha}$
case also the effective impurity band has a mixed X$np$ and C$2p$ character.
Here $n=2$ for X=B,N and $n=3$ for X=Al,P. In the case of X=$\cal V$ the picture
is even simpler. The metallic band in the middle of the diamond gap
well isolated from both bands is almost entirely due to the
nearest neighbor C atoms surrounding the vacant site.
In this work we present our preliminary result on comparison of the
impurity band formation for the above elements for various
doping rates $\alpha=1/128,1/54,1/16$.

\section{Method of calculation}
In this study we used the Plane Wave-Pseudopotential
Quantum-ESPRESSO code~\cite{pwscf}. We used Density Functional Theory with 
General Gradient Approximation (GGA). The GGA exchange-correlation functional 
which has been used is PBE~\cite{PBE}. 
We employed ultrasoft pseudopotential~\cite{Vanderbilt} to describe electron-ion interaction. 
The energy cutoff for expansion of wave function in plane wave was 25 Ry, 
and 150 Ry has 
been used for expansion of charge density. We chose 
$2 \times 2 \times 2$ , $3 \times 3 \times 3$ and $4 \times 4 \times 4$ 
supercell with one vacancy or  with a defect, 
that is X$_{\alpha}$C$_{1-\alpha}$ with  $\alpha=1/128,1/54,1/16$.  
B, N, Al, P are chosen to substitutionally replace one carbon atom 
in a diamond structure. The k-point is sampled according to table~\ref{tab1}.
To accelerate electronic structure calculation, we use
Methfessel and Paxton’s Fermi-level smearing method (width 0.01 Ry)~\cite{smearing}. 
The impurity has been only substituted by one of carbon atom in diamond structure 
without any relaxation.  

\begin{table}[ht]
\caption{\label{tab1} The k-point sampling for different supercells.}
\begin{ruledtabular}
\begin{tabular}{lcccc}
Supercell & k-point grid &   number of           \\
          &              & atomic sites          \\
\hline
    $2\times 2 \times 2$        &   $6\times 6 \times 6$ & 16 \\ 
    $3\times 3 \times 3$        &   $4\times 4 \times 4$ & 54  \\ 
    $4\times 4 \times 4$        &   $3\times 3 \times 3$ & 128  \\ 
\end{tabular}
\end{ruledtabular}
\end{table}

\begin{figure}[th]
\vspace{-11.0 cm}
\hspace{-06.0 mm}
\includegraphics[width=3.6in]{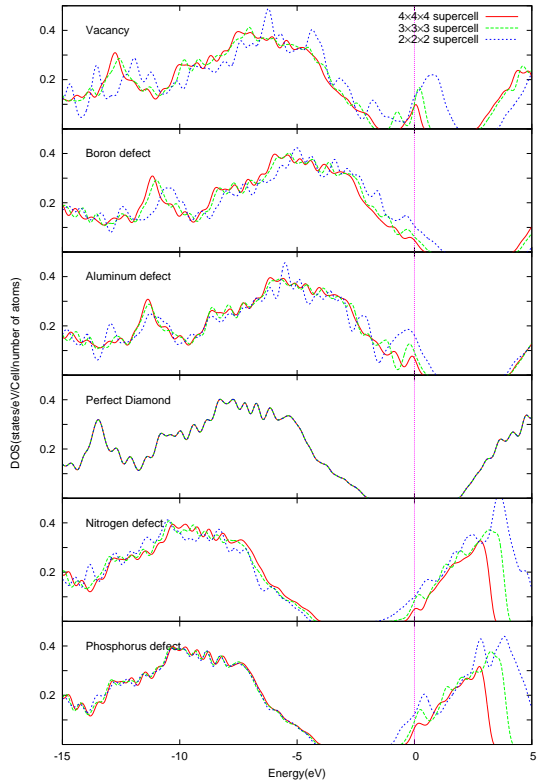}
\caption{\label{fig:dos} The density of state (DOS) for various concentrations of
B, Al, N, P and $\cal V$ impurities in diamond.}
\end{figure}

\section{Results and discussion}
The results of this work has been summarized in Fig.~\ref{fig:dos}
where we have plotted the density of state (DOS) for various 
impurities and different impurity concentrations.
We have shifted the data such that the Fermi level in all cases is
at $E_F=0$. The impurity band formation for the vacancy and Aluminum  
are the most clear among the cases studied here. Their bandwidths 
are between 1.5 eV to 0.5 eV for different impurity concentration. 

   For the perfect diamond lattice the DOS for all sizes of the unit cell
coincide and give the clean diamond gap. However substitutional doping
with B and Al starts to create acceptor levels on top of the valence band
which broaden into bands in the regime of few percent impurity concentration.
Also in this regime the relevant orbitals of the impurity atoms will hybridize
with the C$2p$ orbitals, and the impurity band has a mixed character, similar
to the case of Zhang-Rice singlet of cuprates~\cite{ZRS}. It is clearly seen 
that the impurity band peak in the case of Al is stronger than B. Also the 
value of the DOS at the Fermi level ($\rho_0$) is larger for Al doping than
B doping.
   N and P similarly create donor levels at the buttom of the conduction
band. Again qualitatively one can see that P tends to give a sharper 
impurity band peak than the N. Here also the impurity band arises from a
combination of nearest neighbor C$2p$ and N$2p$ or P$3p$ bands.

   In the case of vacancy, the story is different. 
First of all, since $\cal V$ is neither
acceptor, nor donor level, the resulting band will be in the 
middle of the gap, well isolated from the valence and conduction band.
Secondly the impurity band arises from the n.n. carbon atoms surrounding the
vacancy. This qualitative picture can be inferred by looking at the 
orbital and site resolved partial DOS (not shown here).
Therefore for the effective one band model of the impurities proposed
by Baskaran~\cite{Baskaran}, doping by $\cal V$ seems to be 
more suitable than the other cases studied in this investigation.
  
   Note that in this study we have ignored issues like the formation 
of various complexes. For example, 
nitrogen doping in diamond usually leads to the formation of
nitrogen-vacancy complex~\cite{MainwoodReview}. Ignoring such 
complications, one can use the argument of Mott 
to get a rough estimate of the typical density needed 
to make the the resulting half-filled impurity band superconduct: 
The critical concentration needed for metalization is given by
$a_B^3 n_c=1/4$, where the Bohr radius $a_B$ of the impurity 
 can be estimated from the binding energy 
$E_B$ of the impurity levels as $a_B=\frac{e^2}{2\varepsilon_0 E_B}$.
For typically $0.5-1.5$ eV binding energies of say, Al, $\cal V$, 
the Bohr radius will be $1-3$ atomic units, giving a critical 
concentration on the scale of $\sim 10^{21}$ cm$^{-1}$ or $5-15$ percents.
For the concentrations affordable in our calculations, at 
$\alpha=1/16\approx 6\%$ non of the impurities studied here
fills in the gap. However Cu doping at nearly $6\%$ already metalizes
the diamond~\cite{CuC}.

 According to a disordered RVB mechanism suggested by
Baskaran~\cite{Baskaran}, for such an effective single band at 
half filling the critical temperature for the superconductivity
is given by $k_{B} T_{c}\approx \frac{W}{2} e^{\frac{-1}{\rho_0 J}}$, 
where J is superexchange, $W$ the band width, and $\rho_0$ is 
the DOS at the Fermi level obtained from the DFT bands.
The effective one band model must have large enough $W$ to escape
the Mott-Hubbard splitting in the large $U$ limit, such that 
the half-filled band picture remains valid. Also the disordered
nature of the impurity centers will start to localize the states 
close to the impurity band edges. Again the bandwidth $W$ must be
wide enough such that the mobility edge will not cross the Fermi
level.

   If one assumes that the superconducting mechanism
remains the same for various impurities X studied here, vacancy
and Al doping offer a more clear
picture of the half-filled band undergoing Anderson-Mott to 
RVB superconducting scenario of Baskaran~\cite{Baskaran}.  
The enhancement of $\rho_0$
observed for the case of similar concentration of Al and $\cal V$ 
is advantageous in giving larger transition temperature than the
case of Boron. Note that larger $\rho_0$ even within the BCS picture
is an advantage of $\cal V$ and Al doping compared to doping by B.

   In terms of practical fabrication, utilizing the heavily vacant
diamond may offer a new method in addition to high temperature high
pressure techniques used in production of CVD diamond doped with various
elements. Irradiation by different particles may offer a method of
producing high concentration of vacancies in diamond which can
possibly lead to higher transition temperature than the CVD 
diamond doped with elements.

\section{acknowledgments}
S.A.J. was supproted by Alavi Group Ltd.
S.A.J. would likes to thank S. Maekawa for the hospitality during his visit
to the institute for materials research, Tohoku university.
We are grateful to G. Baskaran for useful comments and discussion.


\end{document}